\documentstyle[aps,prl,amstex,epsfig]{revtex}

\begin{document}
\title{Strong-Coupling Features Due to
Quasiparticle Interactions
in Two Dimensional
Superconductors}

\author{D. Coffey}
\address{Department of Physics, State University of New York,
Buffalo, New York 14260}
\date{\today}

\maketitle
\begin{abstract}
I calculate the effect of interactions among
superconducting quasiparticles in two-dimensional(2D)
a superconductor at T=0.
The strength of the effective interaction
among the quasiparticles
is essentially given by the screened Coulomb
interaction
which has strength at low frequency
because of the gapless nature of the plasmon.
This is in contrast to three dimensions where the effective
interaction has negligible weight at frequencies
$\sim \Delta$, the superconducting gap.
The quasiparticle interactions give
rise to strong-coupling effects in
experimental quantities which are beyond the conventional Eliashberg 
treatment of superconductivity.  The present calculation offers 
an explanation of why these effects 
are much larger in 2D than in 3D superconductors
and, in particular, why the analogous
 strong-coupling effects due to quasiparticle interactions
are seen in data on the quasi-2D cuprate superconductors.
the strong-coupling features seen in data on the cuprates
are discussed in light of the present calculation.
\end{abstract}
\pacs{PACS numbers: 74.50.+r, 74.40.+k}

\section {Introduction}
The BCS theory of superconductivity and 
its extension Eliashberg theory describe the 
electronic properties of the superconducting state
in terms of  self-energies which arise from the interaction
of the superconducting quasiparticles with a boson
which provides the mechanism for superconductivity\cite{rev1,rev2}.
It is assumed that the direct Coulomb interaction
among the quasiparticles leads to renormalizations of the 
quasiparticle effective mass and lifetime 
in the normal state but that these 
renormalizations are unaffected by the superconducting transition.
Within this approximation direct interactions among the
superconducting quasiparticles are ignored and the only indication
of interactions in the superconducting state are the
$\alpha^2F(\omega)$ strong-coupling features
which appear in tunneling experiments and measurements of
optical conductivity.
The $\alpha^2F(\omega)$ function then gives information on the 
density of the bosons and the strength of interaction with
the quasiparticles.
This has proved to be a very successful treatment of
the properties of conventional superconductors
where this mean field approximation
for the direct long-range Coulomb interaction
is justified by data.
Models for the mechanism of high T$_c$ superconductivity
are based on the magnetic properties arising form strong
short range Coulomb correlations as described in the Hubbard model.
Treatments of these models also frequently assume that
the normalizations determined in the normal state are unaffected
by the transition to superconductivity.

However quasiparticle interactions in the ordered state
have been shown to renormalize the superconducting gap,
change the temperature dependence of the gap,
and lead to finite quasiparticle
lifetimes\cite{bishop,coffey90,coffey93a,coffey93b,monthoux}.
In the cuprate superconductors
quasiparticle interactions
lead to a feature in tunneling conductance across an
superconductor-insulator-superconductor($g_{SIS}$) junction
at $3\Delta$ in addition to
the peak at $2\Delta$ expected from the
mean field approximation\cite{Zas,Zas97a}.
It is the scaling of the dip feature with $\Delta$
which distinguishes these corrections from the usual 
features in $\alpha^2F(\omega)$
associated with the electron-phonon interaction.
This scaling was originally demonstrated by comparing 
the tunneling conductance curves for a set of
cuprate superconductors
whose T$_c$'s varied from 5K to 100K\cite{Zas}.
More recently the intrinsic nature of this strong-coupling feature
was demonstrated by comparing the tunneling conductance 
for 
Bi$_2$Sr$_2$CaCu$_2$O$_{8+\delta}$ as a function of doping
and seeing the dip feature scale with the position of the 
2$\Delta$ peak\cite{Zas97a}.
 The size of the dip feature
is much larger than the conventional electron-phonon 
corrections in the cuprates.
This dip feature has been seen in the tunneling results of a
number of groups\cite{Wnuk,Mandrus,Chen,Hartge,Ved}
and the corresponding feature in the spectral density has also
been identified in ARPES(Angle Resolved Photoemission Spectroscopy)
data on Bi$_2$Sr$_2$CaCu$_2$O$_{8+\delta}$\cite{Hwu,Dessau,Shen}
and has recently generated a renewed interest\cite{Shen97,Norman97}.
 
The magnitude of these corrections to the mean field
treatment of superconductivity in the cuprates distinguishes the 
cuprates from conventional superconductors.
Corrections to the mean field approximation, although present
in conventional superconductors, do not in general
lead to experimentally detected features.
In this paper I argue that it is the low-dimensionality of the
cuprate superconductors which is responsible for
the comparatively large effects of
quasiparticle interactions in the cuprates compared to
conventional 3D superconductors.
In low dimensional systems fluctuations lead to the
destruction of long-range order.
In one dimension quantum fluctuations of the order parameter
are sufficient to destroy long-range order\cite{Efetov}
and in two-dimensions
thermal fluctuations destroy the order\cite{Rice,Hohenberg}.
Given this the effects of quantum fluctuations
on superconducting properties
would be expected to
be more important in 2D than in 3D.
This is investigated
using a model Hamiltonian
for 2D and 3D s-wave superconductors at zero temperature
which includes the Coulomb interaction between electrons
and the pairing interaction originally introduced by BCS which
is characterized by a magnitude $V_0$ for electrons within an energy
$\omega_D$ of the Fermi surface.

In this paper I show how the long-range nature of the Coulomb interaction
leads to a
qualitative difference in the quasiparticle properties
of
2D and 3D s-wave superconductors.
As Anderson\cite{PWA} pointed out the long-range nature of the
Coulomb interaction leads to the disappearance of the gapless
mode associated with the phase of the superconductor order
parameter in three dimensions and collective mode becomes the 
plasmon at energies much higher the energy scale of 
superconductivity.
As a result the collective mode has no influence on superconducting
properties\cite{Rick,Bardasis,Prange,Martin}.
 This is not the case  in two dimensions
where the collective mode remains gapless
just as the plasmon mode does for the 2D
electron gas\cite{Dunmore,DSarma,Hwang,cote,Griffin,vdMarel95a}.
This leads to the possibility for the collective mode
having a strong influence on quasiparticle properties unlike the
case in three dimensions.
Interactions among electrons are 
described by the Coulomb interaction and the pairing interaction
responsible for the s-wave superconductivity in the
generalized
random phase approximation. 
I show that the quasiparticle interactions in the 2D
superconductor are stronger than in the 3D case
and that this difference can be traced to the low lying collective mode
and the enhanced magnitude of the effective interaction at low 
frequencies compared to that in 3D.

Most models of the electronic properties of the cuprate
superconductors have emphasized their layered nature
and have employed 2D or
quasi-2D Hamiltonians.
These models have stressed short-range correlations
either through an on-site Hubbard repulsion, in the
weak or strong-coupling limits,
or phenomenologically through residual short-range antiferromagnetic
order which survives in the doped materials\cite{Scal96,Pines97}.
In these models a tight-binding bandstructure is
used with either an on-site Hubbard repulsion coming from
 the weak-coupling side or the $t-J$ model coming from the
strong-coupling side.
Both approaches have been shown to be consistent with the
$d_{x^2-y^2}$ symmetry of the order parameter.
The importance of long-range correlations
coming from the Coulomb interaction in quasi-2D systems,
such as the layered cuprate superconductors, have not been considered
in these treatments but potentially have a strong influence on the
properties of quasi-2D superconductors.
 
The single-particle self-energies are calculated 
here for 2D and 3D
s-wave superconductors
and it is found that their magnitude
 in 2D is roughly ten times the value in 3D.
The single-particle self-energies also have a stronger frequency
dependence in 2D than in 3D
which leads to stronger signatures of quasiparticle interactions
in the spectral density and 
in the tunneling conductance of 2D superconductors compared to
those in 3D superconductors.
I calculate the strong-coupling corrections in model s-wave
superconductor introduced and discuss how these results are related to
experimental data on the cuprates.
The paper is organized as follows. In section II the model is introduced.
In section III the nature of the collective modes  and
their coupling to quasiparticle excitations are reviewed.
The calculated single-particle self-energies are discussed in
section IV and the influence of quasiparticle
interactions on 
measured quantities is discussed in section V.
The implications of the results presented here for work on the
properties of the cuprates are discussed in section VI.
A preliminary report of this work has appeared 
elsewhere\cite{coffey97}.
\section {Model}\vskip 8pt
The starting point of the present analysis is a Hamiltonian, Eq.(1),
describing a system of fermions interacting via a potential,
$U(\vec q)$.
\begin{equation}
H = \sum_{\vec k, \sigma} \xi_{\vec k} c^{\dagger}_{\vec k \sigma}
c_{\vec k \sigma}  
+
{{1}\over{2}}\sum_{\vec k\prime,\vec k,\vec q,\sigma}
 V(\vec k, \vec q) c^{\dagger}_{\vec k - \vec q \sigma}
c^{\dagger}_{-\vec k\prime + \vec q -\sigma}
     c_{-\vec k\prime -\sigma} c_{\vec k \sigma}
+{{1}\over{2}}\sum_{\vec k\prime,\vec k,\vec q, \alpha, \beta}
U(\vec q) c^{\dagger}_{\vec k - \vec q \alpha}
c^{\dagger}_{-\vec k\prime + \vec q \beta}
     c_{-\vec k\prime \beta} c_{\vec k \alpha}
\end{equation}
where $ \xi_{\vec k}= {{|\vec k|^2}\over{2m}}
-{{ k_F^2}\over{2m}}$,
$k_F$ is the Fermi momentum, 
$U(\vec q)={{2\pi e^2}\over{\epsilon_s|\vec q|}}$,
$\epsilon_s$
is the static dielectric constant of the material
and $V(\vec k, \vec q)$ is the same 
pairing interaction as introduced by BCS,
 $V(\vec k, \vec q)=-V_0\Theta(w_D-|\xi_{\vec k}|)
\Theta(w_D-|\xi_{\vec k -\vec q}|)$,
with
$\Theta(x)=0$ for $x<0$ and $=1$ for $x>0$.
$w_D$ is a cutoff energy for the pairing interaction, which is 
a free parameter in the current calculation, was taken to be the 
Debye energy by BCS.
Assuming that the ground state of the system at low temperatures is
an s-wave superconductor,
there is a transformation of the Hamiltonian to
one written in terms of the quasiparticle operators,
$\hat \gamma_{\vec k \sigma}$,
which destroy this state.
Transforming the Hamiltonian it is written in terms 
of these quasiparticle operators, $\hat \gamma^{\dagger}
_{\vec k \sigma}$
and $\hat \gamma _{\vec k \sigma}$,
the quasiparticle spectrum,
$E_{\vec p}=\sqrt{\xi_{\vec p}+\Delta^2}$,
 and the coherence factors,
$u^2_{\vec p}=1-v^2_{\vec p}
={{1}\over{2}}(1+{{\xi_{\vec p}}\over{E_{\vec p}}})$. 
$\Delta$ is determined by the weak
coupling gap equation.
The Hamiltonian becomes
\begin{equation}
H = \sum_{\vec k \sigma} E_{\vec k} \gamma^{\dagger}_{\vec k \sigma}
\gamma_{\vec k \sigma} 
+H^{(R)}_{C}+H^{(L)}_{C}+H^{(3)}_{C}+H^{(4)}_{C}
+H^{(phase)}_{P}+H^{(amp)}_{P}+H^{(3)}_{P}.
\end{equation}
The terms in the Hamiltonian from the Coulomb interaction are,
\begin{equation}
H^{(R)}_{C}=
\sum_{\vec k_1,\vec k_2,\vec q} 
U(\vec q)m(\vec k_1+\vec q,\vec k_1)m(\vec k_2+\vec q,\vec k_2)
\gamma^{\dagger}_{\vec k_1+\vec q \uparrow}
\gamma^{\dagger}_{-\vec k_1 \downarrow}
\gamma_{-\vec k_2 \downarrow}\gamma_{\vec k_2+\vec q \uparrow},
\end{equation}
\begin{equation}
H^{(L)}_{C}=   
{{1}\over{2}}\sum_{\vec k_1,\vec k_2,\vec Q,\sigma,\sigma'}
U(\vec k_1-\vec k_2)n(\vec Q-\vec k_1,\vec Q- \vec k_2)
n(\vec k_1,\vec k_2)
\gamma^{\dagger}_{\vec Q -\vec k_1\sigma}    
\gamma^{\dagger}_{\vec k_1 \sigma'}
\gamma_{\vec k_2 \sigma'} 
\gamma_{\vec Q-\vec k_2 \sigma} 
,
\end{equation}
\begin{equation}
H^{(3)}_{C}=   
\sum_{\vec k_1,\vec k_2,\vec q,\sigma}
U(\vec q)n(\vec k_1+\vec q,\vec k_1)m(\vec k_2-\vec q,\vec k_2)
\gamma^{\dagger}_{\vec k_1-\vec q \sigma}
\gamma^{\dagger}_{\vec k_2+\vec q \uparrow}
\gamma^{\dagger}_{-\vec k_2 \downarrow} 
\gamma_{\vec k_1 \sigma} 
+h.c.,
\end{equation}
\begin{equation}
H^{(4)}_{C}=   
{{1}\over{2}}\sum_{\vec k_1,\vec k_2,\vec q}
U(\vec q)m(\vec k_1+\vec q,\vec k_1)m(\vec k_2-\vec q,\vec k_2)
\gamma^{\dagger}_{\vec k_1+\vec q \uparrow}
\gamma^{\dagger}_{-\vec k_1 \downarrow}
\gamma^{\dagger}_{\vec k_2-\vec q \uparrow} 
\gamma^{\dagger}_{-\vec k_2 \downarrow} 
+h.c.
\end{equation}
In the $H^{(R)}_{C}$ 
the net momentum of the two quasiparticles is the same
as that in the Coulomb interaction. 
The $H^{(L)}_{C}$ term describes the scattering of pairs of 
quasiparticles whose net momentum $\vec Q$ is different from that in the
Coulomb interaction and appears in the calculation of 
ladder diagrams with the Coulomb interaction.
The $H^{(4)}_{C}$ term describes the creation or destruction
of two pairs of quasiparticles one with momentum $\vec q$ and the other
with momentum $-\vec q$ in which $\vec q$ is also the momentum carried
in the Coulomb interaction.
The functions $l(\vec k_1,\vec k_2)$,  $m(\vec k_1,\vec k_2)$,
 $n(\vec k_1,\vec k_2)$ and  $p(\vec k_1,\vec k_2)$
are the usual combinations of coherence factors\cite{JRSbook},
\begin{eqnarray}
m(\vec k+\vec q,\vec k)&=&
u_{\vec k+\vec q }v_{-\vec k}+v_{\vec k+\vec q }u_{-\vec k},
\\ \nonumber
n(\vec k+\vec q,\vec k)&=&
u_{\vec k+\vec q }u_{\vec k}-v_{\vec k+\vec q }v_{\vec k},
\\ \nonumber
l(\vec k+\vec q,\vec k)&=&
u_{\vec k+\vec q }u_{-\vec k}+v_{\vec k+\vec q }v_{-\vec k},
\\ \nonumber
p(\vec k+\vec q,\vec k)&=&
u_{\vec k+\vec q }v_{-\vec k}-v_{\vec k+\vec q }u_{-\vec k}.
\end{eqnarray}

The terms from the pairing interaction are,
\begin{equation}
H^{(phase)}_{P}=
{{V_0}\over{2}}\sum_{\vec k_1,\vec k_2,\vec q}
l(\vec q +\vec k_1,\vec k_1)l(\vec q+\vec k_2,\vec k_2)
\biggl[\gamma^{\dagger}_{\vec k_1+\vec q \uparrow}
\gamma^{\dagger}_{-\vec k_1 \downarrow}
-\gamma_{-\vec k_2 \downarrow}\gamma_{\vec k_2+\vec q \uparrow}\biggr]
\biggl[\gamma^{\dagger}_{\vec k_1+\vec q \uparrow}
\gamma^{\dagger}_{-\vec k_1 \downarrow}
-\gamma_{-\vec k_2 \downarrow}\gamma_{\vec k_2+\vec q \uparrow}\biggr]
\end{equation}
\begin{equation}
H^{(amp)}_{P}=
{{V_0}\over{2}}\sum_{\vec k_1,\vec k_2,\vec q}
n(\vec q +\vec k_1,\vec k_2)n(\vec q+\vec k_2,\vec k_2)
\biggl[\gamma^{\dagger}_{\vec k_1+\vec q \uparrow}
\gamma^{\dagger}_{-\vec k_1 \downarrow}
+\gamma_{-\vec k_2 \downarrow}\gamma_{\vec k_2+\vec q \uparrow}\biggr]
\biggl[\gamma^{\dagger}_{\vec k_1+\vec q \uparrow}
\gamma^{\dagger}_{-\vec k_1 \downarrow}
+\gamma_{-\vec k_2 \downarrow}\gamma_{\vec k_2+\vec q \uparrow}\biggr]
\end{equation}
\begin{equation}
H^{(2)}_{P}=
{{V_0}\over{2}}\sum_{\vec k_1,\vec k_2,\vec Q,\sigma,\sigma'}
\Biggl[u_{\vec k_1+\vec Q }v_{-\vec k_1}v_{\vec k_2+\vec Q }u_{-\vec k_2}
+v_{\vec k_1+\vec Q }u_{-\vec k_1}u_{\vec k_2+\vec Q }v_{-\vec k_2}\Biggr]
\gamma^{\dagger}_{\vec Q -\vec k_1 \sigma}
\gamma^{\dagger}_{\vec k_1 \sigma'}
\gamma_{\vec k_2 \sigma'}
\gamma_{\vec Q -\vec k_2 \sigma},
\end{equation}
\begin{equation}
H^{(3)}_{P}=   
V_0\sum_{\vec k_1,\vec k_2,\vec q,\sigma}
\Biggl[u_{\vec k+\vec q }u_{\vec k}v_{\vec k+\vec q }u_{-\vec k}-
v_{\vec k+\vec q }u_{-\vec k}u_{\vec k+\vec q }v_{-\vec k}\Biggr]
\gamma^{\dagger}_{\vec k_1-\vec q \sigma}
\gamma^{\dagger}_{\vec k_2 +\vec q \uparrow}
\gamma^{\dagger}_{-\vec k_2 \downarrow} 
\gamma_{\vec k_1 \sigma} 
+h.c.
\end{equation}
The $H^{(phase)}_{P}$ and $H^{(amp)}_{P}$ terms generate phase and
amplitude fluctuations of the order parameter.  

In normal ordering the $\gamma^{\dagger}_{\vec k \sigma}$
and $\gamma_{\vec k \sigma}$ operators
terms
quadratic in the new operators as well as terms without 
operators are generated from the kinetic energy and
interaction terms.
Considering first these terms from the kinetic energy
and pairing interaction they include the effect of the
pairing interaction calculated in the Hartree approximation.
Ignoring the shift in the chemical potential due to the pairing interaction
and using the weak-coupling gap equation these terms 
give the kinetic energy term,
$\sum_{\vec k \sigma} E_{\vec k} \gamma^{\dagger}_{\vec k \sigma}
\gamma_{\vec k \sigma}$,
 in the transformed Hamiltonian.
Consequently in calculating the effect of the interactions in Equation(2)
all contributions which include Hartree contributions to
the quasiparticle lines from the pairing interaction 
should be dropped
to avoid double-counting.

The quadratic terms arising in the normal ordering of the Coulomb
parts of the Hamiltonian lead to the Hartree-Fock spectrum.
In the normal state in 3D this leads to a large shift in the 
chemical potential, $\delta \mu = -{{e^2 k_F}\over {\pi}}$,
and a logarithmically diverging effective mass\cite{Mahan}.
In 2D the self-energy leading to the Hartree-Fock spectrum is
$\Sigma^X_{2D}(k)={{-2e^2k_F}\over{\pi}}E({{k}\over{k_F}})$
for $k < k_F$ and
$\Sigma^X_{2D}(k)={{-2e^2k_F}\over{\pi}}[E({{k_F}\over{k}})
-(1-({{k_F}\over{k}})^2)K({{k_F}\over{k}})]$ for $k > k_F$, 
where $K(x)$ and $E(x)$ are elliptic integrals of the first and second kind.
As in 3D there is a large shift in the chemical potential,
 $\delta \mu = -{{2e^2 k_F}\over {\pi}}$
and the effective mass diverges logarithmically.
These results for the Coulomb interaction in 
the Hartree Fock approximation are strongly effected by higher
order terms in the perturbation expansion and the 
long range interactions should be screened\cite{Khodel,Nozieres92}.
The statically screened Coulomb interaction leads 
to a smaller shift in the
chemical potential and to a  modest reduction in the effective mass.
In the calculations discussed in the present paper
the normal state quasiparticle spectrum is measured from
the renormalized chemical potential and the reduction in the 
effective mass due to the statically screened Coulomb interaction
is ignored.
The single-particle self-energies for the
$\gamma^{\dagger}_{\vec k \sigma}
\gamma_{\vec k \sigma}$   
and  for the $\gamma^{\dagger}_{\vec k \uparrow}
\gamma^{\dagger}_{-\vec k \downarrow}$
propagators
are calculated at zero temperature
using the generalized random phase approximation(GRPA).
In this approximation the effective interaction is essentially the
screened Coulomb interaction minus its static limit.

\section{The Effective Interaction and Collective Modes}
\subsection{The Effective Interaction}
The single-particle self-energies 
are calculated at zero temperature
using the generalized random phase approximation(GRPA).
In the GRPA
ring diagrams, where each ring is the two quasiparticle propagator,
 are summed in which the
vertices are dressed by the pairing interaction.
In the absence of vertex corrections the
two-quasiparticle propagator is
\begin{equation}
A_{00}(\vec q, \omega)=\sum_{\vec k}
{{1}\over{2}}
(1-{{\xi_{\vec k}\xi_{\vec q-\vec k}}\over{E_{\vec q- \vec k}E_{\vec k}}}
+{{\Delta^2}\over{E_{\vec q- \vec k}E_{\vec k}}})
{{2( E_{\vec q- \vec k}+E_{\vec k}) }\over
{ \omega^2-(E_{\vec q- \vec k}+E_{\vec k})^2}}
\end{equation}
The interaction connecting the dressed two-quasiparticle
propagators carries the net momentum and frequency of the 
two quasiparticles and
is described by terms $H^{(R)}_C$ and $H^{(2)}_P$
in the Hamiltonian.  However the magnitude of the pairing interaction
is negligible compared to the Coulomb interaction 
and so $H^{(2)}_P$
can be ignored.
Scaling energies with respect to $\Delta$
and momenta with respect to $p_F$, the strength of the
interaction 
is given by a dimensionless parameter
${{2\pi e^2}\over{\epsilon_sp_F}}N(0)={{2}\over{\epsilon_sa_Bp_F}}$,
where $a_B$
is the Bohr radius.
The strength of the effective interaction
is determined by the product $\epsilon_sp_F$
which is a free parameter in the model.

At zero temperature
the vertex corrections in the rings are given by the
repeated scattering the two quasiparticles which is described by the 
$H^{(phase)}_P$
and the $H^{(amp)}_P$ terms in the Hamiltonian.
Calculating a "rung" in this ladder sum by summing over the 
intermediate quasiparticle momenta,
one finds 
\begin{eqnarray}
\Biggl(
B_{20}(\vec q, \omega)
l(\vec q &-&\vec k_1,\vec k_1)l(\vec q-\vec k_2,\vec k_2)
+C_{20}(\vec q,\omega)
n(\vec q -\vec k_1,\vec k_1)n(\vec q-\vec k_2,\vec k_2)
\\ \nonumber
&+&E_{20}(\vec q, \omega)
\Biggl[l(\vec q -\vec k_1,\vec k_2)n(\vec q-\vec k_2,\vec k_2)
+l(\vec q -\vec k_1,\vec k_1)n(\vec q-\vec k_2,\vec k_2)\Biggr]
\Biggr)
\gamma^{\dagger}_{\vec q+\vec k_1 \uparrow}
\gamma^{\dagger}_{\vec k_1 \downarrow}
\gamma_{\vec q-\vec k_2 \uparrow}
\gamma_{\vec k_2 \downarrow}
\end{eqnarray}
where
\begin{eqnarray}
B_{20}(\vec q, \omega)&=&\sum_{\vec k}
{{1}\over{2}}
(1+{{\xi_{\vec k}\xi_{\vec q-\vec k}}\over{E_{\vec q- \vec k}E_{\vec k}}}
+{{\Delta^2}\over{E_{\vec q- \vec k}E_{\vec k}}})
{{E_{\vec q- \vec k}+E_{\vec k}}\over
{\omega^2-(E_{\vec q- \vec k}+E_{\vec k})^2}}
\\ \nonumber
C_{20}(\vec q, \omega)&=&\sum_{\vec k}
{{1}\over{2}}
(1+{{\xi_{\vec k}\xi_{\vec q-\vec k}}\over{E_{\vec q- \vec k}E_{\vec k}}}
-{{\Delta^2}\over{E_{\vec q- \vec k}E_{\vec k}}})
{{E_{\vec q- \vec k}+E_{\vec k}}\over
{\omega^2-(E_{\vec q- \vec k}+E_{\vec k})^2}}
\\ \nonumber
E_{20}(\vec q, \omega)&=&\sum_{\vec k}
{{1}\over{2}}
({{\xi_{\vec k}}\over{E_{\vec k}}}
+{{\xi_{\vec q-\vec k}}\over{E_{\vec q- \vec k}}})
{{E_{\vec q- \vec k}+E_{\vec k}}\over
{\omega^2-(E_{\vec q- \vec k}+E_{\vec k})^2}}
\end{eqnarray}
Summing over $\vec k$, $E_{20}(\vec q, \omega)$ vanishes with 
particle-hole symmetry.
$B_{20}(\vec q, \omega)$ and $C_{20}(\vec q, \omega)$
are the phase and amplitude fluctuations of the
order parameter, respectively.
As a result of particle-hole symmetry 
these two channels are decoupled from one another 
and the ladder sum becomes
\begin{equation}
\Biggl[{{V_0}\over{1-V_0B_{20}(\vec q, \omega)}}
l(\vec q -\vec k_1,\vec k_1)l(\vec q-\vec k_2,\vec k_2) 
+{{V_0}\over{1-V_0C_{20}(\vec q,\omega)}} 
n(\vec q -\vec k_1,\vec k_1)n(\vec q-\vec k_2,\vec k_2)\Biggr] 
\gamma^{\dagger}_{\vec q+\vec k_1 \uparrow}
\gamma^{\dagger}_{\vec k_1 \downarrow} 
\gamma_{\vec q-\vec k_2 \uparrow} 
\gamma_{\vec k_2 \downarrow}. 
\end{equation}
This decoupling of amplitude fluctuations
from charge fluctuations has been noted previously
by Wu and Griffin\cite{Griffin}.

The two different combinations of coherence factors in the
phase and amplitude fluctuation channels, $l(\vec q -\vec k_1,\vec k_1)$
and $n(\vec q -\vec k_1,\vec k_1)$, lead to different couplings to the
$H^{(R)}_C$ vertex
which 
are given by a function
$c_{20}(\vec q, \omega)$, for the phase fluctuations,
and by $d_{20}(\vec q, \omega)$, for the amplitude fluctuations,
where
\begin{eqnarray}
c_{20}(\vec q, \omega)&=&\sum_{\vec k}
({{\Delta \omega}\over{E_{\vec q- \vec k}E_{\vec k}}})
{{E_{\vec q- \vec k}+E_{\vec k}}\over
{\omega^2-(E_{\vec q- \vec k}+E_{\vec k})^2}}
\\ \nonumber
d_{20}(\vec q, \omega)&=&\sum_{\vec k}
{{(\xi_{\vec k}+\xi_{\vec q-\vec k})\Delta}\over
{E_{\vec q- \vec k}E_{\vec k}}}
{{E_{\vec q- \vec k}+E_{\vec k}}\over
{\omega^2-(E_{\vec q- \vec k}+E_{\vec k})^2}}
\end{eqnarray}
The coupling of the $H^{(R)}_C$ to amplitude fluctuations of 
the order parameter is zero because of particle-hole symmetry.
As discussed above  the static part of the effective interaction 
leads to a renormalization of the chemical potential and the Hartree-Fock
spectrum and is to be dropped assuming that energies are to be measured from the
renormalized chemical potential.
The effective interaction becomes
\begin{equation}
V_{eff}(\vec q, \omega)=
{{U(\vec q)}\over
{1-U(\vec q)\Pi(\vec q, \omega)}}
-{{U(\vec q)}\over
{1-U(\vec q)P(\vec q, \omega=0)}}
\end{equation}
where
\begin{equation}
\Pi(\vec q, \omega)=A_{00}(\vec q, \omega) -
{{V_0c^2_{20}(\vec q, \omega)}\over{1-V_0B_{20}(\vec q, \omega)}}
\end{equation}
Since the weak coupling gap equation,
which determines the magnitude of $\Delta$ in the present calculation,
 is $1=B_{20}(\vec q=0, \omega=0)$,
$1-V_0B_{20}(\vec q, \omega)$ becomes 
$V_0\delta B_{20}(\vec q, \omega)$,
the $V_o$ factor cancels in Eq. (18),
and 
there is no the explicit dependence of $\Pi(\vec q, \omega)$ on $V_0$.
Furthermore since the energies of interest here are $<10\Delta$,
$\epsilon_sp_F$ is the most important parameter
and $\Delta$ sets the energy scale.
\subsection{Collective Modes}
Within this approximation,
the imaginary part of the effective interaction is
\begin{equation}
V''_{eff}(\vec q, \omega)= 
U^2(q)\chi^{''}(q,\omega)=
{{U^2(\vec q)\Pi''_A(\vec q, \omega)}\over
{|1-U(\vec q)\Pi_A(\vec q, \omega)|^2}}+
U(\vec q)
\delta[1-U(\vec q)\Pi_A(\vec q, \omega)]
\end{equation}
The second term gives the collective mode contribution.
Collective modes
have been investigated extensively for 3D
superconductors\cite{PWA,Rick,Bardasis}
 and more recently for 2D
and quasi-2D
superconductors\cite{DSarma,Hwang,cote,Griffin,vdMarel95a}.
As is well-known the energy of the collective mode in 3D
superconductors is comparable to the plasma energy and 
the collective mode has no influence on properties at energies
$\sim \Delta$.
On the other hand 
$\omega_{\vec q} = \sqrt{ {{2\pi e^2 N(0)}\over{\Delta^2}}q}$
at long wavelengths in 2D superconductors,
where $N(0)=4\pi m/h^{2}$.
As a result $V_{eff}(\vec q, \omega)$ has an imaginary part
given by $\pi Z_{\vec q}\delta(\omega -\omega_{\vec q})$
for values of $\omega \sim \Delta$, where
$Z_{\vec q}=
[U^2(\vec q){{\partial \Pi_A(q,\omega)}
\over {\partial \omega}}]^{-1}|_{\omega=\omega_{\vec q}}$.
\vskip 8pt
$V''_{eff}(\vec q, \omega)$ is $U^2(\vec q)$ times the screened 
response function which is the dynamic
structure factor, $S(q,\omega)$,
calculated in the random phase approximation.
Collective modes exhaust the $f-$sum rule,
$\int d\omega  \omega S(q,\omega) = {{Nq^2}\over{2m}}$, 
at longwavelengths\cite{PinesNoz} so that
\begin{equation}
\int d\omega  \omega S(q,\omega) \simeq Z_q\omega_q= {{Nq^2}\over{2m}}
\end{equation}
 $N$ is the number density of
particles
and the weight in the collective mode is 
$Z_q\simeq N{{q^2}\over{2m}}{{1}\over{\omega_{\vec q}}}$
in the long wavelength limit.
As $\omega_{\vec q} \rightarrow 2\Delta$,  $\omega_{\vec q}$
increases less rapidly than $\sqrt{q}$ and 
is
Landau damped in the two quasiparticle continuum.
This is shown in figure 5 for $\epsilon_sp_F=8\pi\AA^{-1}$. 
At the same time $Z_{\vec q}$ also starts to increase less 
rapidly than $q^2$ and eventually goes to zero as $\omega_{\vec q}$ 
reaches $2\Delta$.

In Figure 1 $f_{q}$ is the fraction of the $f-$sum given by the collective,
$f_{q}= Z_{\vec q}/{{Nq^2}\over{2m}}$.
When $\omega_{\vec q} \rightarrow 2\Delta$,
$f_q \rightarrow 0$.
At larger wavenumbers, 
the collective excitation reemerges as a resonance above the continuum.
In the normal state when the collective mode, the plasmon,
is calculated in the
random phase approximation, the excitation is a pole of the
response function at the same wavenumbers.
  However in the present calculation
the imaginary part of $A_{00}(\vec q, \omega)$ behaves as
$\sim N(0){{\Delta^2}\over{\omega^2}}$ at high frequencies.
This is 
because the $\hat \gamma_{\vec k \sigma}$ operators are mixtures
of $\hat c_{\vec k \sigma}$ and $\hat c^{\dagger}_{\vec k -\sigma}$
and so $A_{00}(\vec q, \omega)$ is a mixture of both normal state
particle-hole and particle-particle propagators.
Consequently there is a resonance at the plasmon energy
rather than a pole.

In Figure 2 I show the plasmon resonance in $\chi^{''}(\vec q, \omega)$
as a function of $\vec q$ for $\epsilon_sp_F=8\pi\AA^{-1}$.
The important point about the effective interaction in two-dimensions is
that there is weight at low frequencies comparable to
$ \Delta$ 
at long wavelengths.
For example with $\epsilon_sp_F=8\pi\AA^{-1}$,
$\omega_{\vec q} < 10\Delta$ for  $|\vec q| < 0.125p_F$.   As a result 
interactions between quasiparticles at low energies $\sim \Delta$
are much stronger than the interactions between the corresponding
quasiparticles in 3D superconductors.
As the value of $\epsilon_sp_F$ decreases the strength of the 
interaction increases  and
the resonance will emerge from the continuum
at smaller $q$ and $\omega$. I will discuss the dependence of the
results on $\epsilon_sp_F$ below.

\noindent
\begin{figure}[t]
\unitlength1cm
\begin{minipage}[t]{15.0cm}
{\centerline{\epsfxsize=150mm
\epsfysize=100mm
\epsffile{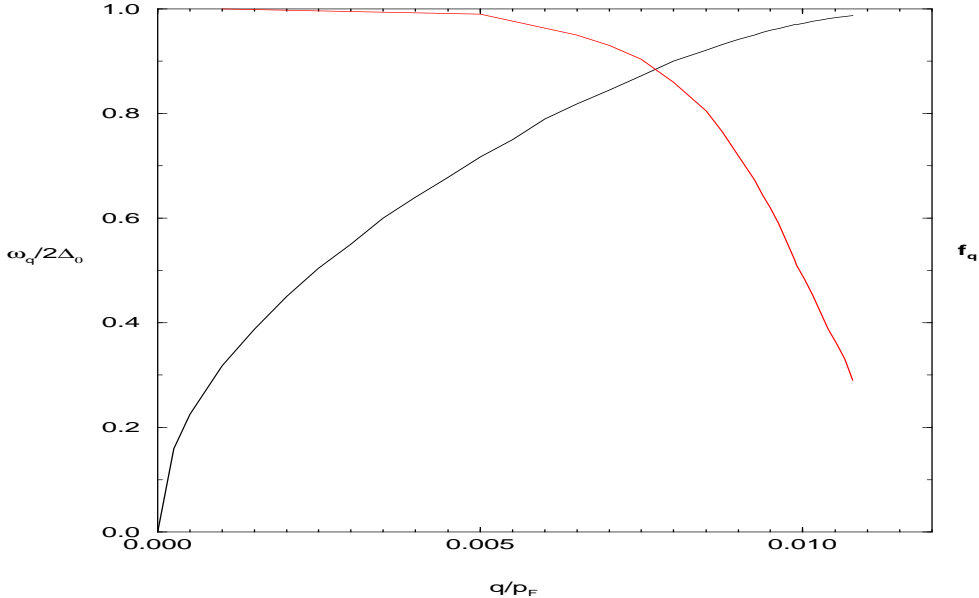}}}
\vskip0.1cm
\caption{Collective mode
and weight in f-sum
 for $\Delta=0.0273e_F$.
and $\epsilon_s p_F=8\pi\AA^{-1}$.
}
\label{fig1}
\end{minipage}
\hfill
\end{figure}
 
\noindent
\begin{figure}[t]
\unitlength1cm
\begin{minipage}[t]{15.0cm}
{\centerline{\epsfxsize=150mm
\epsfysize=100mm
\epsffile{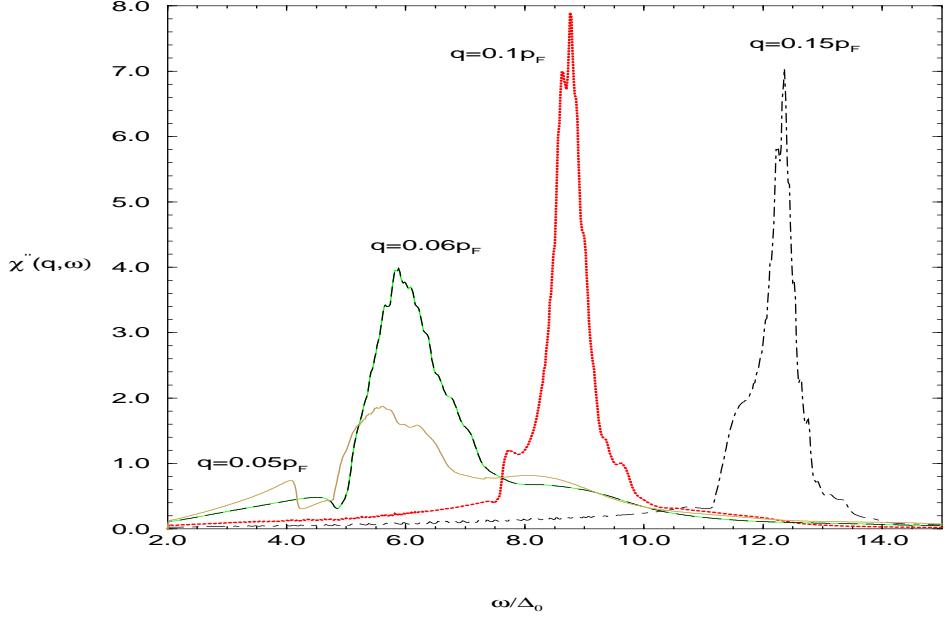}}}
\vskip0.1cm
\caption{Collective mode
emerging from the continuum
seen as a resonance in the imaginary part of the response function,
$\chi^{''}(q,\omega)$.
As $q$ increases the resonance appears at higher energies compared
to $\Delta$ and becomes more like the normal state plasmon.
Eventually it is Landau-damped in the quasiparticle continuum
just the normal state plasmon.}
\label{fig2}
\end{minipage}
\hfill
\end{figure}
 
Fertig and DasSarma\cite{DSarma}
 have calculated collective modes in
layered superconductors allowing for hopping
between planes. They found that collective modes exist in the 
gap for a limited range of momenta in the layers, $|\vec q|$,
and transverse to the layers, $q_z$.
This region of $|\vec q|$ and $q_z$ values
shrinks with increasing strength of the interlayer hopping, $W$.
Interlayer hopping also leads to a gap in the collective mode
dispersion at long wavelengths
comparable in magnitude to $W$.
The important feature of the collective mode spectrum
from the point of view of the present work is that they are
at energies of order $\Delta$.
As long as this is the case low energy quasiparticle interactions will be
enhanced above those in 3D superconductors.
Estimates given by Fertig and DasSarma suggest that this is the case 
in the cuprates.
A detailed account of quasiparticle interactions in 
layered superconductors is in preparation\cite{Coffey-lsc}.

\vskip 8pt
 
\section { Single-Particle Self-Energies} 
The single-particle self-energy for the
$\gamma^{\dagger}_{\vec k_1 \sigma}
\gamma_{\vec k_1 \sigma}$
and $\gamma_{-\vec k_1 \downarrow}
\gamma_{\vec k_1 \uparrow}$
propagators
are calculated at zero temperature
using the generalized random phase approximation(GRPA).
The self-energies associated with the two single-particle
propagators,
$G_{\gamma^{\dagger}\gamma}(\vec p,iE_n)$
and $G_{\gamma\gamma}(\vec p,iE_n)$
are,
\begin{eqnarray}
\Sigma_{\gamma^{\dagger}\gamma}(\vec p,iE_n)&=&
-{{1}\over{2}}\sum_{\vec q} \Biggl [
tanh({{E_{\vec p -\vec q}}\over{2T}})
[n^2(\vec p- \vec q, \vec p)V_e(\vec q, iE_n-E_{\vec p -\vec q})
-
m^2(\vec p- \vec q, \vec p)
V_e(\vec q, -iE_n-E_{-\vec p-\vec q})]
\\ \nonumber &+&
\int^{\infty}_{-\infty}{{d\omega }\over{\pi }}
coth({{\omega}\over{2T}})V''_e(\vec q, \omega)
\biggl [
{{n^2(\vec p -\vec q, \vec p)}
\over{iE_n - \omega -E_{\vec p -\vec q} }}
+{{m^2(\vec p+\vec q ,-\vec p)}
\over{iE_n + \omega +E_{-\vec p -\vec q} }}\biggr
] \Biggr]
\end{eqnarray}
\begin{eqnarray}
\Sigma_{\gamma\gamma}(\vec p,iE_n)&=&
-{{1}\over{2}}\sum_{\vec q}
\Biggl [tanh({{E_{\vec p -\vec q}}\over{2T}})
m(\vec p, \vec p-\vec q)n(\vec p, \vec p-\vec q)
\times
\biggl [V_e(\vec q, iE_n-E_{\vec p -\vec q})
-V_e(\vec q, -iE_n-E_{\vec p -\vec q})\biggr]
\\ \nonumber &-&
\int^{\infty}_{-\infty}{{d\omega}\over{\pi }}
coth({{\omega}\over{2T}}) V''_e(\vec q, \omega)
{{ \omega +E_{\vec p -\vec q}}
\over{E_n^2 + (\omega +E_{\vec p -\vec q})^2 }}
\biggr ]\end{eqnarray}

Analytically continuing to the real axis
and taking the zero temperature limit,
the
real
part of
the retarded self-energies are
\begin{eqnarray}
\Sigma'_{\gamma^{\dagger}\gamma}(\vec p,E)=
-{{1}\over{2}}\sum_{\vec q}\biggl[n^2(\vec p- \vec q, \vec p)&\Biggl (&
V'_e(\vec q, -E-E_{\vec p -\vec q})
+\int^{\infty}_0{{d\omega}\over{\pi}}
{{2(E -E_{\vec p -\vec q})V''_e(\vec q, \omega)}
\over{(E -E_{\vec p -\vec q})^2-\omega^2 }}\biggr )
\\ \nonumber
&-&m^2(\vec p- \vec q, \vec p)
\Biggl (V'_e(\vec q, E+E_{-\vec p-\vec q})
 +\int^{\infty}_0{{d\omega}\over{\pi}}
{{2(E +E_{-\vec p -\vec q})V''_e(\vec q, \omega)}
\over{(E +E_{-\vec p -\vec q})^2-\omega^2 }}
 \biggr)
\biggr]
\end{eqnarray}
\begin{eqnarray}
&\Sigma&'_{\gamma\gamma}(\vec p,E)=
-{{1}\over{2}}\sum_{\vec q} 
m(\vec p, \vec p-\vec q)n(\vec p, \vec p-\vec q)
\\ \nonumber
&\times&
\biggl (V'_e(\vec q, E-E_{\vec p -\vec q})
-V'_e(\vec q, -E-E_{\vec p -\vec q})
+\int^{\infty}_0{{d\omega}\over{\pi}}
\biggl [ 
{{2(E -E_{\vec p -\vec q}) V''_e(\vec q, \omega)}
\over{(E -E_{\vec p -\vec q})^2-\omega^2 }}
+{{2(E +E_{-\vec p -\vec q}) V''_e(\vec q, \omega)}
\over{(E +E_{-\vec p -\vec q})^2-\omega^2 }}\biggr
]
\biggr )
\end{eqnarray}
The imaginary parts of the self-energies are,
\begin{eqnarray}
\Sigma''_{\gamma^{\dagger}\gamma}(\vec p,E)=
\sum_{\vec q} \Theta(E-E_{\vec p -\vec q})
{{1}\over{2}}
(1+{{\xi_{\vec p}\xi_{\vec p-\vec q}-\Delta^2}\over
{E_{\vec p- \vec q}E_{\vec p}}})
V''_{eff}(\vec q, E-E_{\vec p -\vec q})
\\ \nonumber
\Sigma''_{\gamma\gamma}(\vec p,E)=
\sum_{\vec q} \Theta(E-E_{\vec p -\vec q})
{{(\xi_{\vec p}+\xi_{\vec p-\vec q})\Delta}\over
{E_{\vec p- \vec q}E_{\vec p}}}
V''_{eff}(\vec q, E-E_{\vec p -\vec q}),
\end{eqnarray}
where
$V'_{eff}(\vec q, \omega)$ and $ V''_{eff}(\vec q, \omega)$
are the real and imaginary parts of the effective interaction.

The calculated imaginary part of 
$\Sigma_{\gamma^{\dagger}\gamma}(\vec p,E)$
is  shown for a two-dimensional s-wave superconductor
in Figure 3 with $\epsilon_sp_F$ equal to 8$\pi\AA^{-1}$,

\noindent
\begin{figure}[t]
\unitlength1cm
\begin{minipage}[t]{15.0cm}
{\centerline{\epsfxsize=150mm
\epsfysize=100mm
\epsffile{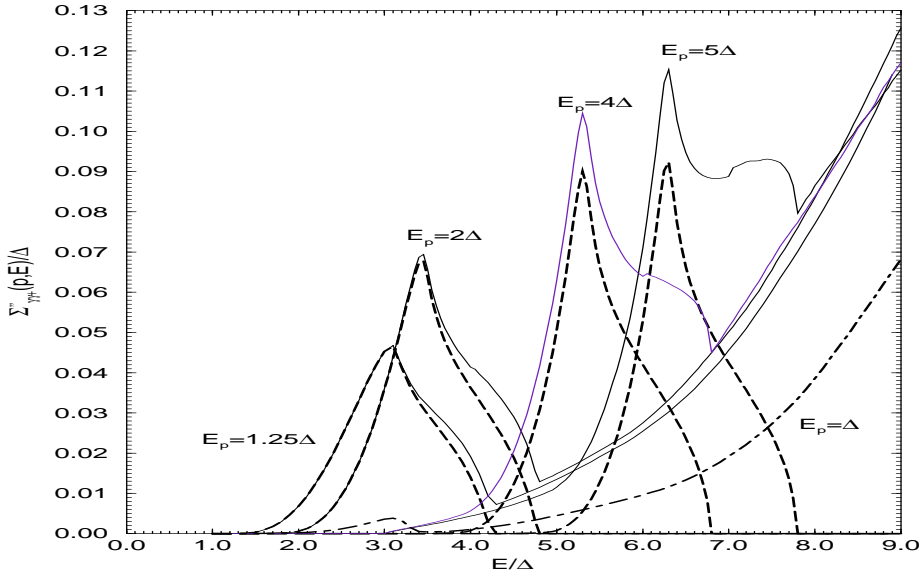}}}
\vskip0.1cm
\caption{Imaginary part of
$\Sigma_{\gamma^{\dagger} \gamma}(\vec p,E)$
for a 2D s-wave superconductor for different values of $E_p$ with
$\epsilon_s p_F=25.13\AA^{-1}$, $\omega_D=0.1e_F$
and
$\Delta=0.0273e_F$.
The dashed line is the contribution from the collective mode
for each $E_p$.
This is very small
for $E_{\vec p}=\Delta$(dot-dash curve).}
\label{fig3}
\end{minipage}
\hfill
\end{figure}

In Figure 3 the dashed curve is the contribution due to scattering off
the
collective mode
and the solid line
 is the sum of the collective mode and continuum
contributions.
In 3D $\Sigma''_{\gamma^{\dagger}\gamma}(\vec p,E)$
increases monotonically with increasing $E>3\Delta_0$
as the phase for decay into quasiparticle pairs increases.
Comparing the 2D and 3D results one notes that
the collective mode contribution is absent from the 3D case
and that $\Sigma''_{\gamma^{\dagger}\gamma}(\vec p,E)$
is an order of magnitude larger in 2D.
There is little dependence on $|\vec p|$ for $|\vec p|> 1.01p_F$
in the continuum contribution to $\Sigma''_{\gamma^{\dagger}\gamma}(\vec p,E)$. This contribution is a monotonically increasing function of energy at these
low energies.
The different coherence factors which appear in
the expression for $\Sigma''_{\gamma\gamma}(\vec p,E)$
lead a smaller magnitude than for
$\Sigma''_{\gamma^{\dagger}\gamma}(\vec p,E)$.
However the energy dependence in the collective and
continuum contributions
are unchanged.

\noindent
\begin{figure}[t]
\unitlength1cm
\begin{minipage}[t]{15.0cm}
{\centerline{\epsfxsize=150mm
\epsfysize=100mm
\epsffile{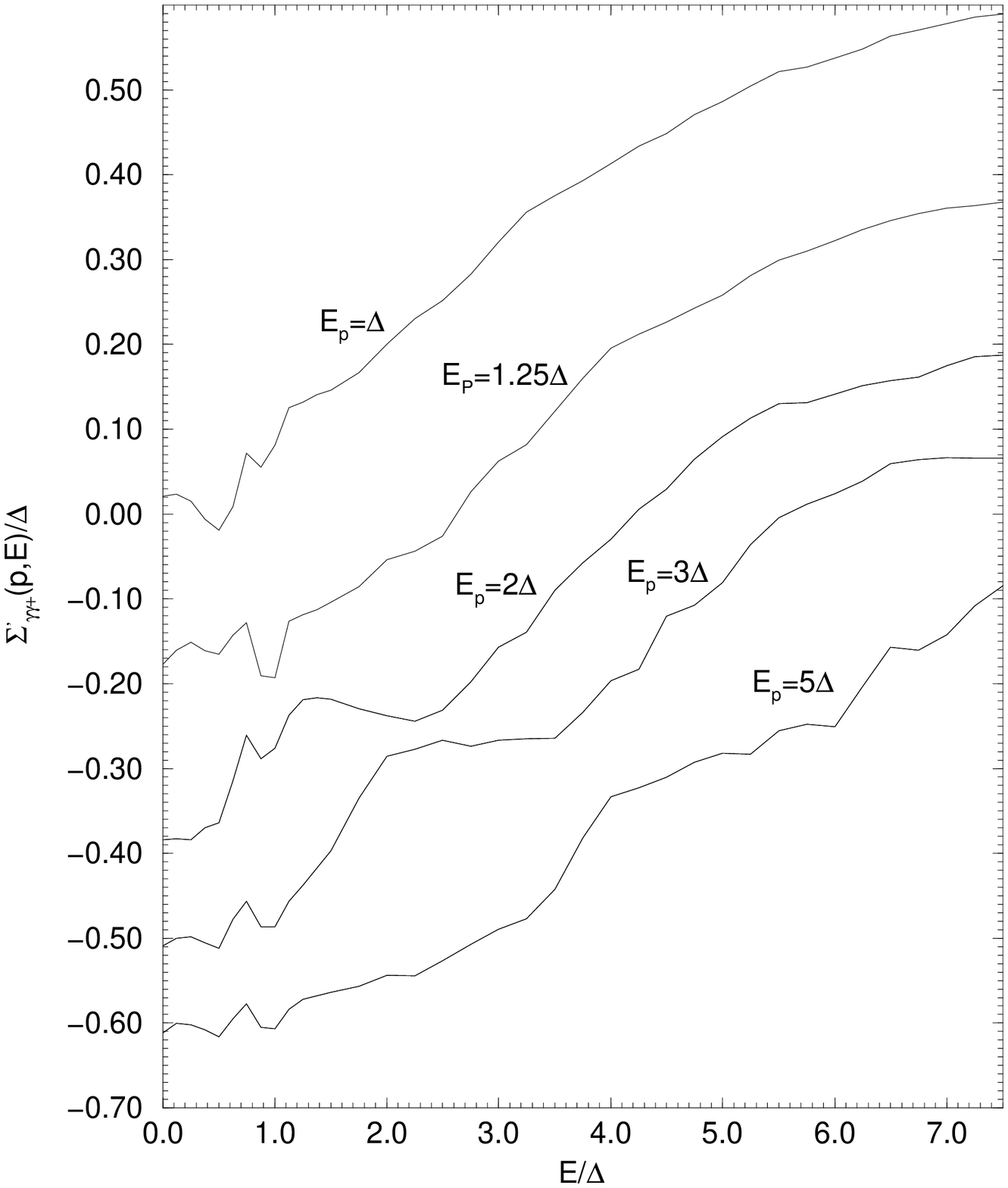}}}
\vskip0.1cm
\caption{Real part of the self-energy,
$\Sigma_{\gamma^{\dagger} \gamma}(\vec p,E)$,
for a 2D s-wave superconductor for different momenta with
$\epsilon_s p_F=25.13\AA^{-1}$, $\omega_D=0.1e_F$
and 
$\Delta=0.0273e_F$.}
\label{fig4}
\end{minipage}
\hfill
\end{figure}

The real part of the $\Sigma_{\gamma^{\dagger}\gamma}(\vec p,E)$
self-energy
is plotted for $\epsilon_sp_F =  8\pi\AA^{-1}$ in Figure 4 and 
for $\epsilon_sp_F =  4\pi\AA^{-1}$ in Figure 5.
From Figure 4 one see that the $E$ dependence is different for
$E < E_p$ where the dependence is non-monotonic, 
for $E \simeq E_p$ where there is very little dependence on $E$
 and for
$E > E_p$, after a rapid increase, there is monotonic almost
linear increase
with $E$.
For $E \simeq 2\Delta$ this rapid increase has a step-like dependence.

In Figure 5 $\Sigma_{\gamma^{\dagger}\gamma}(\vec p,E)$
is seen to have the same form for different values of
$\epsilon_sp_F$ but that the magnitude increases as
$\epsilon_sp_F$ decreases, that is as the magnitude of the
Coulomb interaction increases.
For $E_p > 4\Delta$ the on-shell self-energy
$\Sigma_{\gamma^{\dagger}\gamma}(\vec p,E=E_p)$
is only very weakly dependent on E
and the magnitude is larger for smaller values of
 $\epsilon_sp_F$.
For  $\epsilon_sp_F=8\pi\AA^{-1}$
$\Sigma'_{\gamma^{\dagger}\gamma}(\vec p,E_p) \rightarrow -0.3\Delta$,and for   $\epsilon_sp_F=4\pi\AA^{-1}$
$\Sigma'_{\gamma^{\dagger}\gamma}(\vec p,E_p) \rightarrow -0.78\Delta$,
as $E \rightarrow 10\Delta$.
For values of  $\epsilon_sp_F$
such that $\Sigma'_{\gamma^{\dagger}\gamma}(\vec p,E_p) \simeq E_p$
for
small $E_p$ and
corrections to GRPA should be calculated.
A detailed discussion of the influence of the form of
$\Sigma_{\gamma^{\dagger}\gamma}(\vec p,E)$
on the density of states and the tunneling conductance
is
given in the next section.
The other self-energy, $\Sigma'_{\gamma\gamma}(\vec p,E)$,
 is shown for 2D in
figures 6.
The magnitude of this self-energy is smaller than
the magnitude of
$\Sigma'_{\gamma^{\dagger}\gamma}(\vec p,E_p) \simeq E_p$
and falls off monotonically for large $E$.
Its magnitude also increases with decreasing $\epsilon_sp_F$
but its dependence on $E$ remains unchanged.

Comparing the 2D and 3D cases again one finds that the
magnitude of the real part of the self-energies is roughly
an order of magnitude smaller in 3D than in 2D
and for $E > \Delta_0$ the energy dependence is almost linear
without the step feature of the 2D case.
The same free parameter, $\epsilon_sp_F$,
 is present in the 3D calculations
however because of the magnitude of the self-energies in 3D
${{\Sigma}\over{\Delta}}$ remains small for much smaller values of
$\epsilon_sp_F$.

\noindent
\begin{figure}[t]
\unitlength1cm
\begin{minipage}[t]{15.0cm}
{\centerline{\epsfxsize=150mm
\epsfysize=100mm
\epsffile{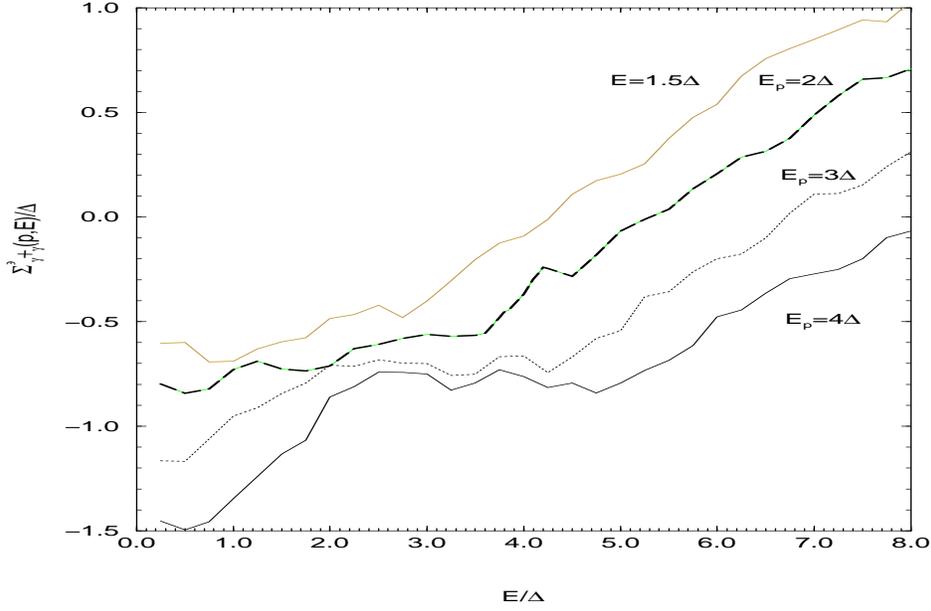}}}
\vskip0.1cm
\caption{Real part of the self-energy,
$\Sigma_{\gamma^{\dagger} \gamma}(\vec p,E)$,
for a 2D s-wave superconductor for different momenta with
$\omega_D=0.1e_F$
and
$\Delta=0.0273e_F$.
for $\epsilon_s p_F=4\pi\AA^{-1}$.
The dependence on E is the same as that for
$\epsilon_s p_F=8\pi\AA^{-1}$ but the magnitude is three times larger.}
\label{fig5}
\end{minipage}
\hfill
\end{figure}

\noindent
\begin{figure}[t]
\unitlength1cm
\begin{minipage}[t]{15.0cm}
{\centerline{\epsfxsize=150mm
\epsfysize=100mm
\epsffile{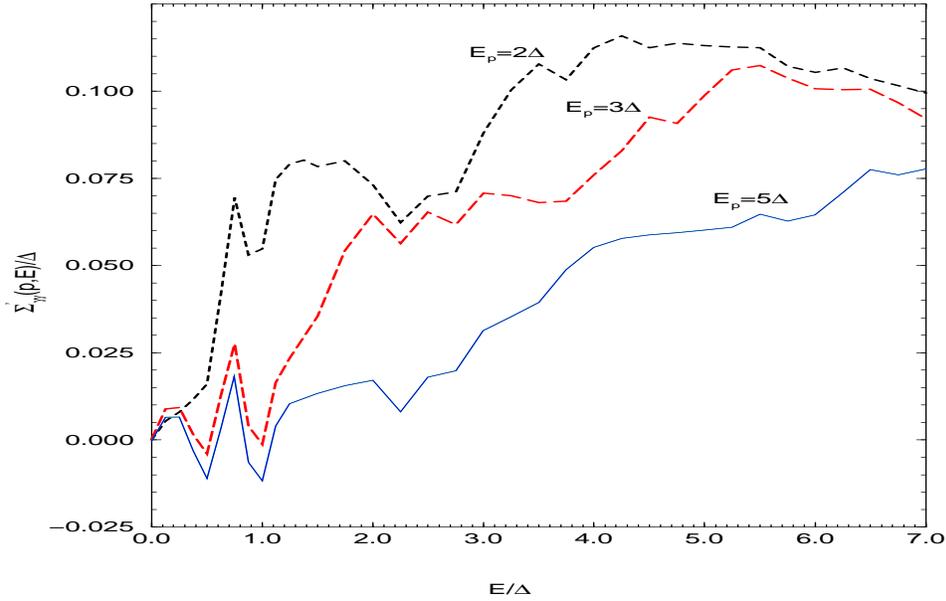}}}
\vskip0.1cm
\caption{Real part of the self-energy,
$\Sigma_{\gamma \gamma}(\vec p,E)$,
for a 2D s-wave superconductor for different momenta with
$\epsilon_s p_F=25.13\AA^{-1}$, $\omega_D=0.1e_F$
and
$\Delta=0.0273e_F$.}
\label{fig6}
\end{minipage}
\hfill
\end{figure}
The same calculation can be carried out for a 2D d-wave
superconductor.  In many models of superconductivity of the cuprates
the d-wave symmetry arises from a pairing
interaction with strong short-range correlations
which are not present in this work.
In a more complete treatment of the charge carriers in the
cuprates both long range and short range parts of the Coulomb
interaction would be included.
As Hwang and Das Sarma\cite{DSarma} have pointed out
the long range Coulomb interaction
leads to quantitatively the same collective mode behavior
in both s- and d- wave superconductors
so that there is the same kind of enhancement of quasiparticle interactions
in a 2D d-wave superconductor as appears in the 2D s-wave case.
In calculations of superconducting properties for
a d-wave superconductor with a tight-binding bandstructure
bandstructure and gap anisotropy
lead to anisotropy in the self-energy\cite{coffey93b}.
Nodes in the gap function also lead to damping of collective modes.
These effects will not lead to a qualitatively different conclusion
from the simple model discussed here
but are clearly important for comparison with data on the
cuprates.

\section{Strong Coupling Features in Tunneling Conductances}
Considering the expression for the the tunneling conductance 
across an superconductor-insulator-superconductor 
junction, 
\begin{equation}
g={{\partial I(V)}\over{\partial V}}
={{\partial}\over{\partial V}} \sum_{\vec k_1 \vec k_2 }{\int^{eV}_{0}} 
d\omega A(\vec k_1, \omega)
A(\vec k_2, eV-\omega) |T|^2
\end{equation}
where $I(V)$ is the current across an SIS junction
and
$T$ is an unspecified matrix element between states on
opposite sides of the junction which in principle 
depends on $\vec k_1$ and $\vec k_2$.
In  this calculation the tunneling
matrix elements are
taken to be constants\cite{Cohen}.
In describing tunneling in systems with strong bandstructure effects
this approximation for the tunneling matrix element is not
always justified\cite{lcoffey97}.
Summing the carrier spins the spectral density $A(\vec k, \omega)$ is 
\begin{equation}
A(\vec k, E)= 
{{2}\over{\pi}}
\Biggl[{{u^2_{\vec k} \Sigma ''_{\gamma^{\dagger}\gamma}(\vec k,E)}\over 
{(E-E_{\vec k}-\Sigma'_{\gamma^{\dagger}\gamma}(\vec k,E))^2
+(\Sigma''_{\gamma^{\dagger}\gamma}(\vec k,E))^2}}
+{{v^2_{\vec k}\Sigma''_{\gamma^{\dagger}\gamma}(-\vec k,-E)}\over
{(E+E_{\vec k}+\Sigma'_{\gamma^{\dagger}\gamma}(-\vec k,-E))^2
+(\Sigma''_{\gamma^{\dagger}\gamma}(-\vec k,-E))^2}}
\Biggr]\end{equation}
 
\noindent
\begin{figure}[t]
\unitlength1cm
\begin{minipage}[t]{15.0cm}
{\centerline{\epsfxsize=150mm
\epsfysize=100mm
\epsffile{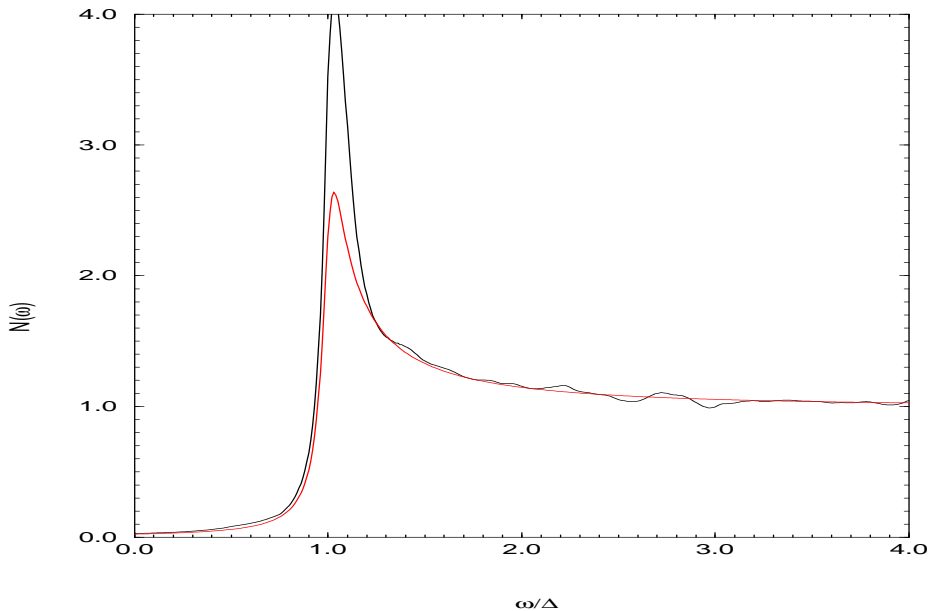}}}
\caption{Comparison between $N(\epsilon)$
with the mean field approximation, $N^{MF}(\epsilon)$
for 2D superconductors.
$N^{MF}(\epsilon)$ was calculated using a Lorentzian with
$\Gamma_0=\Delta/20$.}
\label{fig7(a)}
\end{minipage}
\hfill
\end{figure}

\noindent
\begin{figure}[t]
\unitlength1cm
\begin{minipage}[t]{15.0cm}
{\centerline{\epsfxsize=150mm
\epsfysize=100mm
\epsffile{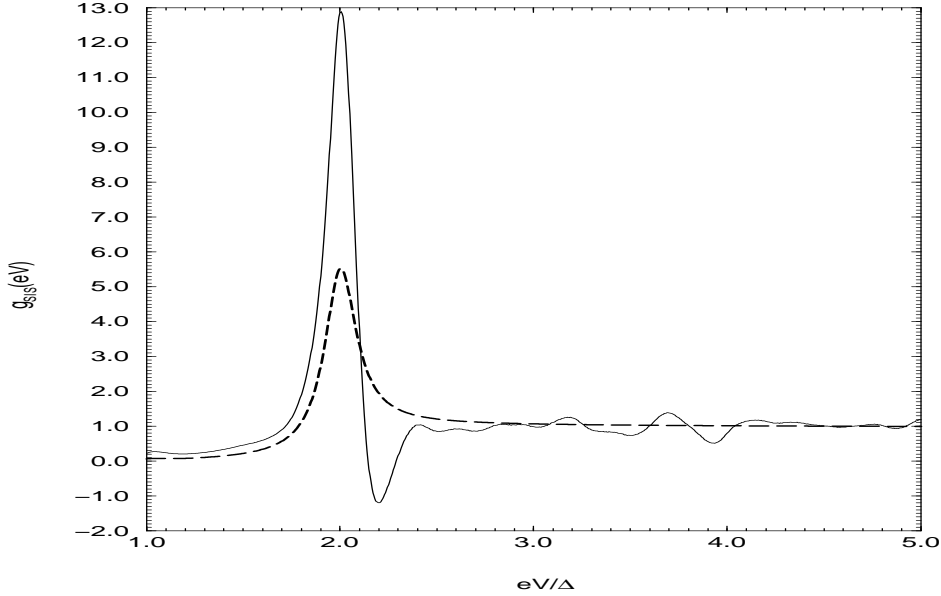}}}
\caption{Comparison between
$g_{SIS}$ corresponding to the $N(\omega)$ in Figure 7(a)
with the mean field approximation
$g^{MF}_{SIS}$(dashed curves),
for 2D superconductors.
$N^{MF}(\epsilon)$ was calculated using a Lorentzian with
$\Gamma_0=\Delta/20$.}
\label{fig7(b)}
\end{minipage}
\hfill
\end{figure}

The tunneling density of states,
essentially the tunneling conductance 
across a superconductor-insulator-normal metal(SIN) junction,
 and the tunneling conductance
 are compared with the mean field approximation in Figure 7
and in Figure 8 for $\epsilon_sp_F=8\pi\AA^{-1}$.
There is a piling up of states at E's $\sim  \Delta$ due to
$\Sigma'_{\gamma^{\dagger}\gamma}(\vec p,E)$
compared to the density of states calculated in the
mean field approximation, 
$N^{MF}(\omega)=\sum_{\vec k}{{\Gamma_0}
\over{(\omega-E_{\vec k})^2+\Gamma_0^2}}$.
In these calculations an extra damping term,
$\Gamma_0={{\Delta_0}\over{20}}$, is added to
$\Sigma''_{\gamma^{\dagger}\gamma}(\vec p,E)$
to mimic the effect of impurity scattering on these features.
 This piling up of states leads to a non-monotonic dependence of the
current across an S-I-S junction as a function of applied voltage
which in turn  results the tunneling conductance shown in Figure 7.
The large variation in $g_{SIS}(eV)$
at $eV \simeq 2\Delta$
comes from the magnitude of
$\Sigma'_{\gamma^{\dagger}\gamma}(\vec p,E)$.
\vskip 8pt
The detailed E-dependence leads to strong-coupling effects
which are seen at different frequencies in $N(\omega)$
and have corresponding features at
frequencies shifted by $\Delta$ in $g_{SIS}(eV)$.
These strong-coupling features are shown in the 
density of states and in the tunneling conductance
in Figures 9 and 10 for different values of
a phenomenological damping factor $\Gamma_0$ in the propagators.
In this way some idea of the sensitivity
of these features with damping can be seen.
The magnitude of these features are at the 5$\%$ level in the density
of states for $\Gamma_0=\Delta/20$.
The most prominent of these features are 
a bump feature between $\sim 2.6\Delta$ and $2.9\Delta$
and the step feature at $\omega =3 \Delta$ in
$N(\omega)$. These lead to  more pronounced features at
$eV=4\Delta$ in $g_{SIS}(eV)$.
This step feature is the analog of the dip feature seen 
in $g_{SIS}(eV)$ at $eV=3\Delta$ in the d-wave cuprate 
superconductors 
whereas the bump feature comes from the collective mode 
contribution as will be discussed below.
The origin of these features is the
self-energy arising from quasiparticle interactions mediated by the
screened Coulomb interaction and not from the pairing interaction.

The on-shell imaginary part of the self-energy,
$\Sigma''_{\gamma^{\dagger}\gamma}(\vec p,E_p)$,
is given to a good approximation for $\epsilon_sp_F=8\pi\AA^{-1}$
by
\begin{equation}
\Sigma''_{\gamma^{\dagger}\gamma}(\vec p,E_p)=
\Gamma_0+0.055\Theta(E_p-3\Delta)[E_p-3\Delta]
\end{equation}
The small magnitude means that the
spectral density at any momentum is to a very good
approximation  a Lorentzian and
there is no structure from the energy
dependence of $\Sigma_{\gamma^{\dagger}\gamma}(\vec p,E)$
at values of E where the spectral density, $A(p,E)$,
has any appreciable magnitude.
If $\Sigma''_{\gamma^{\dagger}\gamma}(\vec p,E)$ is set equal to
$\Gamma_0={{\Delta}\over{20}}$ and the conductance recalculated
the same features appear in the tunneling conductance
at the same energies.
The only difference is that the new curve lies below the curve
calculated with $\Sigma''_{\gamma^{\dagger}\gamma}(\vec p,E)$.
So the $E$ dependence of $\Sigma''_{\gamma^{\dagger}\gamma}(\vec p,E)$
is not responsible for the strong-coupling
 features in the density of
states.
 
\noindent
\begin{figure}[t]
\unitlength1cm
\begin{minipage}[t]{15.0cm}
{\centerline{\epsfxsize=150mm
\epsfysize=100mm
\epsffile{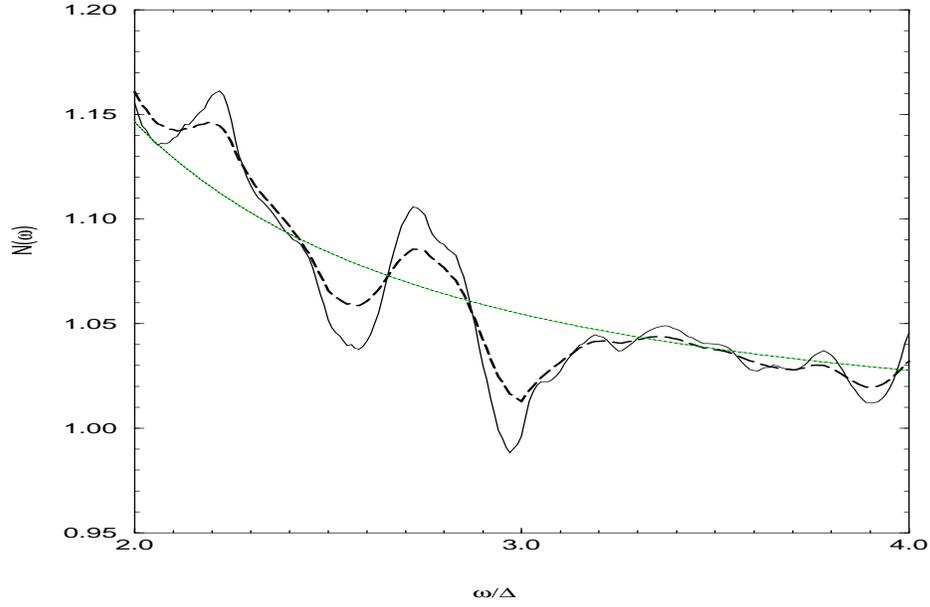}}}
\caption{Strong Coupling effects in the Density of States
shown in Figure 7. The solid curve is the density of states
calculated with $\Gamma_0=\Delta/20$, the heavy dashed curve is calculated
with $\Gamma_0=\Delta/10$ and the dotted
 curve is the mean field density
of states with $\Gamma_0=\Delta/20$.}
\label{fig8a}
\end{minipage}
\hfill
\end{figure}
 
\indent
\begin{figure}[t]
\unitlength1cm
\begin{minipage}[t]{15.0cm}
{\centerline{\epsfxsize=150mm
\epsfysize=100mm
\epsffile{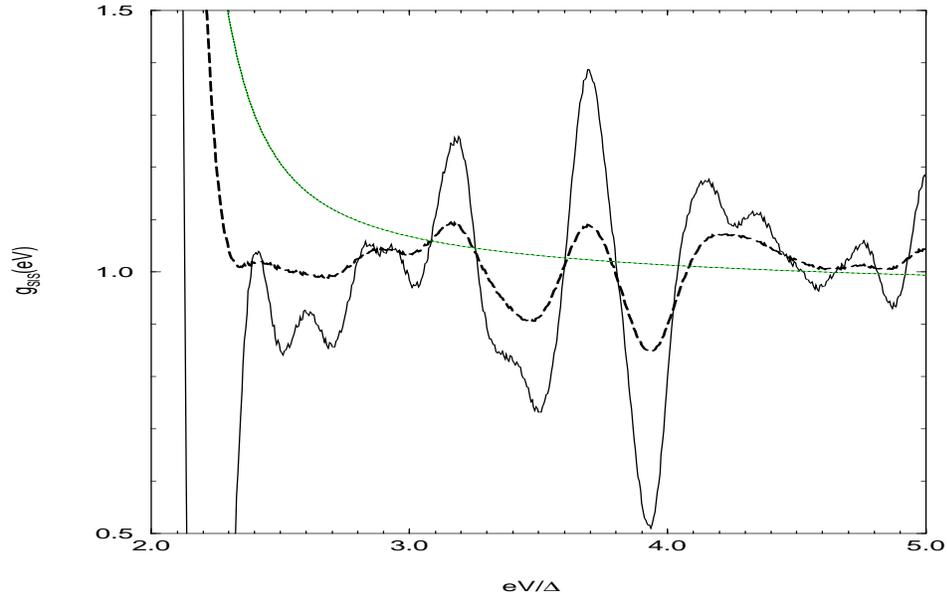}}}
\caption{Strong-coupling effects at $\omega= 4\Delta$
in
$g_{SIS}$
corresponding to that at $\omega = 3\Delta$ in
the density of states shown in Figure 7.
The solid curve is the case of
$\Gamma_0=\Delta/20$, the dashed curve is for $\Gamma_0=\Delta/10$
and the dotted curve is the mean field curve
with $\Gamma_0=\Delta/20$.}
\label{fig8b}
\end{minipage}
\hfill
\end{figure}
 
The real part, $\Sigma'_{\gamma^{\dagger}\gamma}(\vec p,E)$,
can be considered to the sum of two contributions,
the on-shell piece, $\Sigma'_{\gamma^{\dagger}\gamma}(\vec p,E_p)$,
 and the energy dependent part,
$\delta\Sigma'_{\gamma^{\dagger}\gamma}(\vec p,E)$.
The position of the peak in the spectral density
is given by
\begin{equation}
\omega_p=E_p+\Sigma'_{\gamma^{\dagger}\gamma}(\vec p,E_p)
+\delta\Sigma'_{\gamma^{\dagger}\gamma}(\vec p,\omega_p)
\end{equation}
$\omega_p$ is the renormalized quasiparticle spectrum.
$\Sigma'_{\gamma^{\dagger}\gamma}(\vec p,E_p)/\Delta$
is plotted versus $E_P/\Delta$ in Figure 11(curve A).
$\Sigma'_{\gamma^{\dagger}\gamma}(\vec p,E_p)$
has no strong dependence on $E_p$ except in the region
near $E_p=3\Delta$ where the
the features are seen in the density of states.
Beyond $E_p=2\Delta$ it is a slowly increasing function
of $E_p$ except for these fluctuations.
$\omega_p-E_p$ is plotted against $\omega_p$ in the same figure(curve B)
and it is seen to have the same rapid variation near $3\Delta$.
Comparing these curves with the density of states in
 Figure 9 the onset of the bump feature in the
density of states at $\omega\simeq 2.6\Delta$ and the step feature at
$3\Delta$ can be seen to arise from structure of the quasiparticle spectrum.
This structure in $\omega_p$ causes
fluctuations in the density of states of the
renormalized quasiparticle spectrum.
Consequently the strong-coupling features in the
density of states and the SIS tunneling arise from
the momentum and energy dependence of
$\Sigma'_{\gamma^{\dagger}\gamma}(\vec p,E)$
due to quasiparticle interactions.

The form of $\Sigma'_{\gamma^{\dagger}\gamma}(\vec p,\omega_p)$
comes from the continuum and collective contributions to
the imaginary part of the self-energy, 
$\Sigma^{cont''}_{\gamma^{\dagger}\gamma}(\vec p,E)$
and $\Sigma^{coll''}_{\gamma^{\dagger}\gamma}(\vec p,E)$.
From the Kramers-Kronig relation, the energy dependent
real part of the self-energy is
\begin{equation}
\Sigma'_{\gamma^{\dagger}\gamma}(\vec p,\omega_p)=
{{1}\over{\pi}}\int^{\Omega_P}_0
{{\Sigma^{cont''}_{\gamma^{\dagger}\gamma}(\vec p,E')dE'}
\over{\omega_p-E'}}
+{{1}\over{\pi}}\int^{\Omega_P}_0
{{\Sigma^{coll''}_{\gamma^{\dagger}\gamma}(\vec p,E')dE'}
\over{\omega_p-E'}}
+C_p
\end{equation}
where $C_p$ is the integral from the
high energy cutoff,
$\Omega_P$, to infinity and over the negative energies
which give only weak energy dependence.

In order to better understand the from of 
$\Sigma^{cont''}_{\gamma^{\dagger}\gamma}(\vec p,E)$
near $E=3\Delta$,
$\Sigma^{cont''}_{\gamma^{\dagger}\gamma}(\vec p,E)$
and
$\Sigma^{coll''}_{\gamma^{\dagger}\gamma}(\vec p,E)$
can be approximated by simple analytic forms.
From Figure 3 these can be approximate by
$\Sigma^{cont''}_{\gamma^{\dagger}\gamma}(\vec p,E)=
\alpha_p(E-3\Delta)\Theta(E-3\Delta)$ 
with $\alpha_p \simeq 0.055$.
$\Sigma^{cont''}_{\gamma^{\dagger}\gamma}(\vec p,E)$
increases more rapidly than this linear function
by $E \simeq 5\Delta$ but this approximation is sufficient to
give a semi-quantitative account.
I take $\Sigma^{coll''}_{\gamma^{\dagger}\gamma}(\vec p,E)=
\beta_p\Theta(E_p+3\Delta-E)\Theta(E-E_p)f(E-E_p)$,
where $f(E-E_p)$ is a function approximately the form of the 
energy dependence of the collective mode contribution.
Looking at the form of the collective mode contribution in
Figure 3 it can be approximated by 
$f(E-E_p)=({{(E-E_p)}\over{\Delta}})^2\sqrt{{{E_p+3\Delta-E}\over{\Delta}}}$. Taking this form and requiring that the maximum value
of $\Sigma^{coll''}_{\gamma^{\dagger}\gamma}(\vec p,E)$
be $0.09\Delta$ as in Figure 3 requires $\beta_p \simeq 0.02\Delta$
 for $E_p>3\Delta$.
$\beta_p$ grows from zero at
$E_p=\Delta$ and is independent of $E_p$ beyond
$E_p=3\Delta$.

Putting in the forms of the two contributions to the
$\Sigma''_{\gamma^{\dagger}\gamma}(\vec p,E)$
\begin{eqnarray}
\Sigma'_{\gamma^{\dagger}\gamma}(\vec p,E_p)&=& 
{{\alpha_p}\over{\pi}}\int^{\Omega_p}_{3\Delta} 
{{(E'-3\Delta)dE'}
\over{E_p-E'}}
+{{\beta_p}\over{\pi}}\int^{E_p+3\Delta}_{E_p} 
{{f(E')dE'}
\over{E_p-E'}}
+C_p
\cr
&=&
-{{\alpha_p}\over{\pi}}(\Omega_p-3\Delta)
+{{\alpha_p}\over{\pi}}
(E_p-3\Delta)ln{{|E_p-3\Delta|}\over{\Omega_p-3\Delta}}
-{{\beta_p}\over{\pi}}
{{12\sqrt{3}}\over{5}}
+C_p\end{eqnarray}
and
\begin{eqnarray}
\Sigma'_{\gamma^{\dagger}\gamma}(\vec p,\omega_p)=
\Sigma'_{\gamma^{\dagger}\gamma}(\vec p,E_p)
&+&{{\beta_p}\over{\pi}}\biggl[2(3)^{3/2}{{\omega_p-E_p}\over{\Delta}}
+\sqrt{3}\Biggl({{\omega_p-E_p}\over{\Delta}}\Biggr)^2
ln{{|\omega_p-E_p|}\over{3\Delta}}\biggr]
\cr
&+&{{\alpha_p}\over{\pi}}
\Biggl[(\omega_p-3\Delta)ln{{|\omega_p-3\Delta|}\over{\Omega_p-3\Delta}}
-(E_p-3\Delta)ln{{|E_p-3\Delta|}\over{\Omega_p-3\Delta}}\Biggr]
\end{eqnarray}
From Figure 11 
it can be seen that 
$\delta\Sigma'_{\gamma^{\dagger}\gamma}(\vec p,\omega_p)
\simeq 0.01\Delta$
so that
$\omega_p-E_p \simeq \Sigma'_{\gamma^{\dagger}\gamma}(\vec p,E)$
giving 
\begin{eqnarray} 
\Sigma'_{\gamma^{\dagger}\gamma}(\vec p,\omega_p) \simeq 
\Sigma'_{\gamma^{\dagger}\gamma}(\vec p,E_p) 
\Biggl(1+2(3)^{3/2}{{\beta_p}\over{\pi \Delta}}\Biggr)
&+&\sqrt{3}{{\beta_p}\over{\pi}}
\Biggl({{\Sigma'_{\gamma^{\dagger}\gamma}(\vec p,E_p)}\over{\Delta}}\Biggr)^2
ln{{|\Sigma'_{\gamma^{\dagger}\gamma}(\vec p,E_p)|}\over{3\Delta}}
\cr
&+&{{\alpha_p}\over{\pi}}
\Biggl[
(\omega_p-3\Delta)ln{{|\omega_p-3\Delta|}\over{\Omega_P}}
-(E_p-3\Delta)ln{{|E_p-3\Delta|}\over{\Omega_P}}
\Biggr]
\end{eqnarray}

\noindent
\begin{figure}[t]
\unitlength1cm
\begin{minipage}[t]{15.0cm}
{\centerline{\epsfxsize=150mm
\epsfysize=100mm
\epsffile{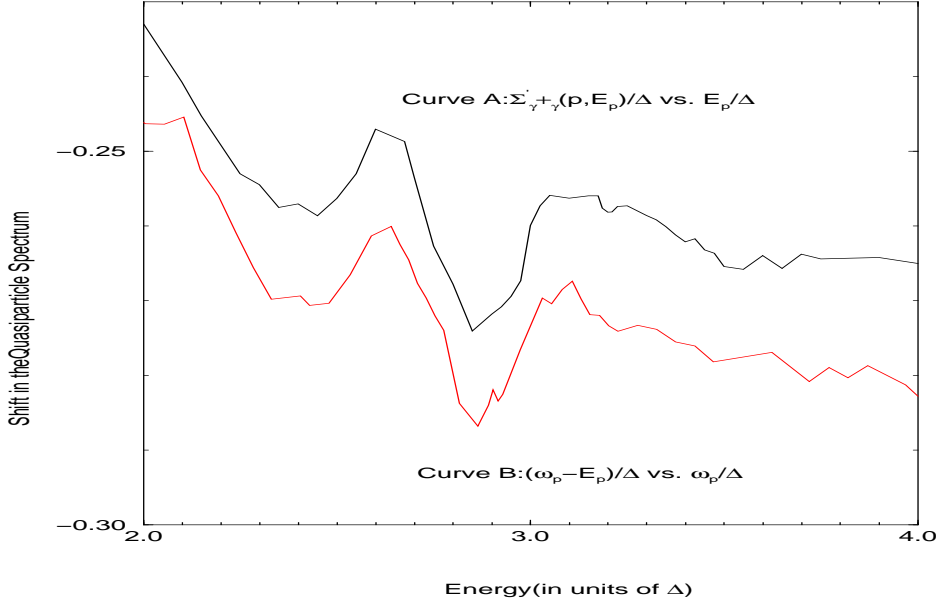}}}
\vskip0.3cm
\caption{The on-shell self-energy,
, as a function of $E_p$ and the shift in the quasiparticle spectrum,
$\omega^*_p$ from $E_p$ as a function of $\omega^*_p$
for
$\epsilon_s p_F=25.13\AA^{-1}$, $\omega_D=0.1e_F$
and
$\Delta=0.0273e_F$.
The position of the features in the quasiparticle spectrum are seen to coincide with the frequencies at which the strong-coupling
 features are seen in the density of states.}
\label{fig11}
\end{minipage}
\hfill
\end{figure}

Using $\beta_p$ discussed above one finds from Equation (33)
that the magnitude of 
$\Sigma'_{\gamma^{\dagger}\gamma}(\vec p,\omega_p)$ is enhanced over that of
$\Sigma'_{\gamma^{\dagger}\gamma}(\vec p,E_p)$
by about 5\%.
Considering the simple form of the functions used this is consistent
with the results shown in Figure 11.
Taking 
 $\Omega_p$
and $\alpha_p$ to be constants 
and allowing 
$\beta_p$ to increase by about ${{1}\over{3}}$
between 
$E_p=2\Delta$ and for $E_p =3\Delta$,
consistent the calculation of
$\Sigma''_{\gamma^{\dagger}\gamma}(\vec p,E)$,
the form of 
$\Sigma'_{\gamma^{\dagger}\gamma}(\vec p,E_p)$ for
$E_p \geq 2.5\Delta$
is seen to come from the combination of the 
${{\alpha_p}\over{\pi}}
(E_p-3\Delta)ln{{|E_p-3\Delta|}\over{\Omega_P}}$
and the ${{\beta_p}\over{\pi}}
{{12\sqrt{3}}\over{5}}$
terms.  
The weak features in curve B, $(\omega_p-E_p)/\Delta$ vs.$\omega_P/\Delta$,
at $\omega_p=3\Delta \pm {{1}\over{2}}\Sigma'_{\gamma^{\dagger}\gamma}(\vec p,E_p)$ come from the term proportional to $\alpha_p$
in Equation (34) where the term in the brackets goes through maxima
in magnitude.
However a better approximation for 
the momentum and energy dependence of
$\Sigma^{''}_{\gamma^{\dagger}\gamma}(\vec p,E)$ is required beyond
the form used here
in order to recover the $E_p$ dependence 
of $\Sigma'_{\gamma^{\dagger}\gamma}(\vec p,E_p)$
between $E_p=2\Delta$ and
$E_p=2.5\Delta$ without a detailed fitting of the 
$C_p$ and $\Omega_p$ functions.
The $\beta_p$ term coming from the collective contribution
to
$\Sigma^{''}_{\gamma^{\dagger}\gamma}(\vec p,E)$ 
is responsible for the bump feature in 
$\Sigma^{'}_{\gamma^{\dagger}\gamma}(\vec p,E_p)$ between
$2.6\Delta < E < 3\Delta$.
It is clear that without the collective mode
contribution the bump in the density of states below
$\omega =3\Delta$ would be absent
and the step at $\omega=3\Delta$
would be more
pronounced.

\section{Discussion}

As well as tunneling conductance and the density of states
discussed here ARPES experiments can provide information
on the single-quasiparticle spectral density once background has 
been taken into account.
In the calculation discussed above the spectral density is a Lorentzian
with no experimentally detectable 
non monotonic energy dependence away from
the peak.  Many-body effects give a non-trivial
quasiparticle spectrum, $\omega_p$,
leading to the strong-coupling effects
discussed above but do not
lead to any signatures in ARPES measurement 
such as the dip feature at $\omega =2\Delta$ in the cuprate
superconductors. A necessary condition that features away from the
quasiparticle peak in the spectral density lead to observable
features in ARPES is that the imaginary
part of the self-energy should have both a strong energy dependence
and be sufficiently large that there is weight away the peak.
In the present case the linear dependence of
$\Sigma''_{\gamma^{\dagger}\gamma}(\vec p,E_p)$ on $E_p$
and the small magnitude combine to ensure that the
energy
dependence of $\Sigma''_{\gamma^{\dagger}\gamma}(\vec p,E)$
leads to no features in ARPES or tunneling conductance.

This is in contrast
to
models used to describe ARPES and tunneling data
in the cuprates\cite{coffey93a,coffey93b,Norman97}
where the strong-coupling 
feature is at $2\Delta$ in the density of
states and the spectral density,
as measured in ARPES,\cite{Shen}
and at $3\Delta$ in the SIS tunneling conductance.
In these models 
a  phenomenological
form for
$\Sigma_{\gamma^{\dagger}\gamma}(\vec p,E)$
is used in which there is a
rapid increase in the magnitude
 of
$\Sigma''_{\gamma^{\dagger}\gamma}(\vec p,E)$
to a value $\sim \Delta$
in some momentum directions
at $E \simeq 2\Delta$
after which any dependence of 
$\Sigma''_{\gamma^{\dagger}\gamma}(\vec p,E)$
on E is weak.
In these models $\Sigma'_{\gamma^{\dagger}\gamma}(\vec p,E)$
is assumed to lead only to a renormalization of the gap.
This model for
$\Sigma''_{\gamma^{\dagger}\gamma}(\vec p,E)$
in the case of a d-wave superconductor
is suggested by Golden-rule calculation of
the
decay of mean field
quasiparticles 
in a tight-binding bandstructure\cite{coffey93a,coffey93b}.
The results of the tight-binding calculations depend on both
the chemical potential
 and the presence of next-nearest neighbor terms
in the bandstructure.  The form of 
$\Sigma''_{\gamma^{\dagger}\gamma}(\vec p,E)$
is consistent with that
arising from low energy antiferromagnetic spin
fluctuations\cite{Monthoux93}. 
The bandstructure and short-range correlations
leading to low energy magnetic excitations
which are responsible for the strong-coupling effects
in the models of the cuprates are absent in the 
calculation reported here.
However the present calculation points to the 
necessity of including long range Coulomb interaction
in order to distinguish low dimensional
from conventional superconductors
and suggests that any model which is
to quantitatively account for these features in the cuprates
should incorporate
the effects of long-range correlations as well as the
short-range correlation associated with the magnetic properties
of the cuprates.
The extent to which a superconductor is low dimensional,
where the effects discussed would be important,l
is determined in the model described here by the weight
in the effective quasiparticle interaction
due to the Coulomb interaction
at low frequencies.
In layered superconductors Das Sarma et al.\cite{DSarma,Hwang}
have shown that depending on interplanar coupling 
and on the direction of the wavevector
the collective mode spectrum,
$\omega_{\vec q}$,
 is at energies $\sim \Delta$ as
$|\vec q| \rightarrow 0$.
 There should still be
substantial weight in the collective mode
at frequencies $< 10\Delta$ in order for quasiparticle interactions
to be enhanced above the 3D case. 
This issue is presently being 
investigated\cite{Coffey-lsc}. 

Since the original identification of the dip feature
in ARPES
there have been suggestions
that it is strongly associated with the mechanism for
superconductivity in the cuprates.
Arnold et al.\cite{Arnold91} identified the strong-coupling features
seen in the data of Dessau et al.\cite{Dessau}.
with a ``bosonic peak" which they proceeded to analyze in terms of the
$\alpha^2F(\omega)$ function of Eliashberg theory.
Analysis of more recent data have lead Shen and Schrieffer\cite{Shen97}
to a model where the quasiparticles are strongly
coupled to a collective mode.
This collective mode contributes to the
line shape in ARPES in the normal state and is strongly influenced by
the superconducting transition. In their picture the dip feature
in the line shape is the consequence of the strong quasiparticle
coupling to the collective mode which provides the pairing
mechanism for hiT$_c$ superconductivity.
In the present case also the collective mode is strongly effected by
the
transition to superconducting ground state
although it is not associated with the mechanism for
superconductivity.
  It decays into
pairs of superconducting quasiparticles once its
 energy reaches $2\Delta$,
reemerges at higher energies and essentially follows the
dispersion of the normal state plasmon. The contribution to 
off-shell self-energy, $\Sigma^{''}(p,E)$,
 from scattering from the plasmon is very weak
at the low energies, $\sim \Delta$,
for the 2D electron gas. 
It leads to a weak feature in the spectral
density even at $p=1.1p_F$ which corresponds to
$E_p \simeq 8\Delta$ for $\Delta=0.0273e_F$
 and would
not be detected easily in experiment\cite{JSK97}. 
\vskip 8pt
\section{Conclusions}
In this paper I have shown how quantum fluctuations in 2D
charged Fermi systems are enhanced over those in 3D.
This result holds for both the normal and superconducting states.
However in the superconducting state they clearly lead to
strong-coupling features in experimental quantities which are
a probe of the superconducting state. 

In the model discussed in this paper
the quasiparticle
interaction responsible for the feature at $\omega=3\Delta$ in
SIN  for s-wave superconductor and the corresponding feature
at $\omega=4\Delta$ in SIS is the Coulomb interaction and
not the pairing interaction giving the s-wave
superconducting ground state.
In the absence of the Coulomb interaction the pairing interaction
would produce the same feature but
would not account for the difference in magnitude of these 
effects
in
conventional 3D superconductors
and the quasi-2D cuprate superconductors.
The anomalously large magnitude of the strong coupling
features seen in the cuprates suggest that the electronic properties 
of these materials are quasi-2D.
This is a basic assumption which most workers in the area have
employed for a long time.
However the long range part of the Coulomb interaction
has been ignored. The present calculation
suggests that, although the interactions
which provide the mechanism for superconductivity in the cuprates
may be short correlations,
 long range Coulomb correlations should also be included.
In particular the pairing mechanism undoubtedly 
contributes to the magnitude of the
feature at $\omega=2\Delta$ in ARPES 
data and to the strong-coupling features seen in
tunneling experiments at $eV=3\Delta$
but they are unlikely to solely account for its
magnitude.

\end{document}